\documentclass[twocolumn,showpacs,preprintnumbers,amsmath,amssymb]{revtex4}

\usepackage{graphicx}
\usepackage{dcolumn}
\usepackage{bm}

\begin{document}

\preprint{APS/123-QED}

\title{Spin-Peierls transition to a Haldane phase}


\author{Hironori Yamaguchi$^{1}$, Hiroki Takahashi$^{1}$, Takashi Kawakami$^{2}$, Kiyomi Okamoto$^{3}$, Toru Sakai$^{3}$, Takeshi Yajima$^{4}$, and Yoshiki Iwasaki$^{5}$}
\affiliation{
$^1$Department of Physical Science, Osaka Metropolitan University, Osaka 599-8531, Japan\\
$^2$Department of Chemistry, Osaka University, Osaka 560-0043, Japan\\
$^3$Graduate School of Science, University of Hyogo, Hyogo 678-1297, Japan\\
$^4$Institute for Solid State Physics, the University of Tokyo, Chiba 277-8581, Japan\\
$^5$Department of Physics, College of Humanities and Sciences, Nihon University, Tokyo 156-8550, Japan
}

Second institution and/or address\\
This line break forced

\date{\today}

\begin{abstract}
We present an organic compound exhibiting spin-Peierls (SP) transition to an effective spin-1 antiferromagnetic uniform chain, that is, the Haldane chain. 
The clear disappearance of magnetization, accompanied by a structural phase transition, is well explained by the deformation to an effective spin-1 Haldane chain.
The flexibility of the molecular orbitals in the organic radical compound allows the transformation of the exchange interactions into the Haldane state with different topologies.
The SP transition in the present compound demonstrates a mechanism different from that of the conventional systems, paving a new path for research in quantum phenomena originating from spin-lattice couplings. 
\end{abstract}

\pacs{75.10.Jm, 
}

\maketitle 
One-dimensional (1D) spin chains, in which localized spins are linearly arranged through exchange interactions, represent the simplest spin model with strong quantum fluctuations.
Various quantum many-body phenomena occur according to the degrees of freedom in 1D spin chains. 
Spin-lattice couplings in spin-1/2 chains give rise to a spin-Peierls (SP) transition to a nonmagnetic quantum phase~\cite{SP1,SP2}, which is the magnetic analog of the Peierls instability of 1D metals~\cite{Peierls}.
The antiferromagnetic (AF) spin-1/2 uniform chain acquires a spin gap through lattice deformation, resulting in an AF alternating chain, revealing that the increase in magnetic energy exceeds the loss of elastic energy due to lattice distortion. 
The SP transition, which corresponds to the boundary between the gapless uniform and gapped alternated states, is a second-order phase transition and has been observed in a variety of 1D materials~\cite{SP_exp1,SP_exp2,SP_exp3,SP_exp4,SP_exp5,SP_exp6}.
Organic materials have particularly many examples because of the lattice softness inherent to molecular-based systems.
In addition, they are mostly unstable upon pressurization, yielding pressure-induced phase transitions to superconducting states via SP ground states~\cite{SPtoSC_01,SPtoSC_02,SPtoSC_03,SPtoSC_04}.


The Haldane state is a well-known example of quantum many-body phenomena in 1D spin models, in which the ground-state topology changes depending on the spin-size~\cite{haldane}.
The Heisenberg AF chain with integer spins demonstrates an energy gap between the nonmagnetic ground state (Haldane state) and the first excited state, whereas that with half-integer spins demonstrates no energy gap. 
The topology of the Haldane state can be described by a valence-bond picture~\cite{AKLT}, in which each integer spin is considered a collection of $S$=1/2 and forms a singlet state between $S$=1/2 spins at the different sites.
In recent research on quantum computers, the application of the edge state of symmetry-protected topological phases in cases with odd-integer spins has been proposed and has attracted attention~\cite{qcom1,qcom2,qcom3}.
The valence-bond picture of the Haldane state can be mapped onto the strong ferromagnetic (F) coupling limit of an F-AF alternating chain with half-integer spin values~\cite{ CuNbO_2,IPACu_masuda,DMA_neutron,NaCuSbO_2,Li3Cu2SbO6}.
The ground-state properties of the spin-1/2 F-AF chains have investigated for various exchange constants ratios, $J_{\rm{F}}/J_{\rm{AF}}$~\cite{hida1,hida2}. 
No discontinuous change in the ground state associated with a phase transition was observed between the Haldane ($|J_{\rm{F}}| \gg J_{\rm{AF}}$) and AF dimer ($|J_{\rm{F}}| \ll  J_{\rm{AF}}$)) states, indicating that the ground state of the spin-1/2 F-AF chain is equivalent to the spin-1 Haldane state.
 

Our material design using verdazyl radicals with diverse molecular structures realizes unconventional spin-1/2 systems, such as the ferromagnetic-leg ladder, quantum pentagon, and random honeycomb, which have not been realized in conventional inorganic materials~\cite{3Cl4FV, a26Cl2V, random}.
The flexibility of the molecular orbital (MO) in the verdazyl radicals enabled us to design spin arrangements composed of intermolecular exchange interactions through molecular design~\cite{fine-tune}.
Moreover, the alternating spin density distribution in $\pi$-conjugated verdazyl systems can readily induce F intermolecular exchange interactions depending on the overlapping molecular orbitals~\cite{F-AF, square_PF6, square_SbF6}.

In this letter, we present a model compound that exhibits an unconventional SP transition.
We synthesized single crystals of the verdazyl-based salt ($p$-MePy-V-$p$-CN)PF$_6$$\cdot$CH$_3$CN [$p$-MePy-V-$p$-CN = 3-(4-methylpyridyl)-1-phenyl-5-(4-cyanophenyl)-verdazyl]. 
Our molecular orbital (MO) calculations and the analysis of the magnetic behavior indicated the SP transition from a spin-1/2 uniform AF chain to a spin-1/2 F-AF alternating chain forming the Haldane state.
Furthermore, we demonstrated that the flexibility of the molecular orbitals in this compound allows the transformation of the exchange interactions into the Haldane state.

We synthesized $p$-MePy-V-$p$-CN using a conventional procedure~\cite{procedure} and prepared an iodide salt of the radical cation ($p$-MePy-V-$p$-CN)I using a reported procedure for salts with similar chemical structures~\cite{mukai}. 
The crystal structures were determined on the basis of intensity data collected using a Rigaku AFC-8R Mercury CCD RA-Micro7 diffractometer and XtaLAB Synergy-S. 
The magnetic susceptibility was measured using a commercial SQUID magnetometer (MPMS-XL, Quantum Design).
The experimental result was corrected for the diamagnetic contributions calculated by Pascal's method.
The specific heat was measured using a commercial calorimeter (PPMS, Quantum Design) by using a thermal relaxation method.
Considering the isotropic nature of organic radical systems, all experiments were performed using small randomly oriented single crystals.
$Ab$ $initio$ MO calculations were performed using the UB3LYP method with the basis set 6-31G and 6-31G($d$,$p$) in the Gaussian 09 program package. 
For the estimation of intermolecular magnetic interaction, we applied our evaluation scheme that have been studied previously~\cite{MOcal}.
The quantum Monte Carlo (QMC) code is based on the directed loop algorithm in the stochastic series expansion representation~\cite{QMC33}. 
The calculations for the spin-1/2 uniform and alternating Heisenberg chains were performed for $N$ = 256 under the periodic boundary condition using the ALPS application~\cite{QMC34,QMC35}.
The numerical diagonalization based on the Lanczos algorithm is performed to obtain the energy eigenvalue and the wave function of the ground state as well as the first excited state of the spin Hamiltonian under the periodic boundary condition up to $N$=28.
For the calculation of the string order parameter Ostr, considering the periodic boundary condition we calculated  $O_{\rm{str}}$($N$/2) with the distance of $N$/2 for the $N$-spin system, and extrapolated them to $N$ ${\rightarrow}$ ${\infty}$.

\begin{figure*}[t]
\begin{center}
\includegraphics[width=40pc]{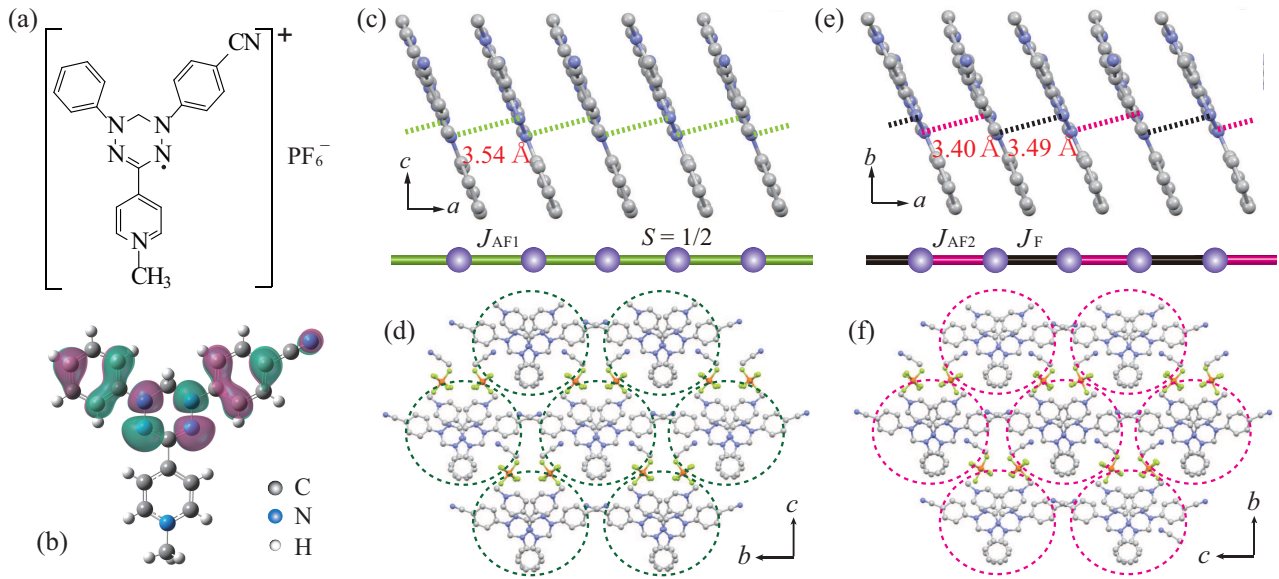}
\caption{(color online) (a) Molecular structure of ($p$-MePy-V-$p$-CN)PF$_6$. 
(b) Singly occupied molecular orbital of $p$-MePy-V-$p$-CN. The purple (green) color indicates isosurfaces of the wave function with positive (negative) sign.
(c) 1D structure forming the spin-1/2 AF uniform chain and (d) interchain structure viewed along the chain direction for $T$ ${\textgreater}$ $T_{\rm{SP}}$.
(e) 1D structure forming the spin-1/2 F-AF alternating chain and (f) interchain structure viewed along the chain direction for $T$ ${\textless}$ $T_{\rm{SP}}$. Hydrogen atoms, PF$_6$ anions, and CH$_3$CN molecules are omitted for clarity. The broken lines indicate N-N short contacts in the molecular pairs associated with $J_{\rm{AF1}}$, $J_{\rm{AF2}}$, and $J_{\rm{F}}$.
The purple spheres represent spin-1/2 in each molecule and are mainly located in the central ring with four N atoms.
The broken circle encloses $p$-MePy-V-$p$-CN radicals comprising each spin chain structure. 
}
\end{center}
\end{figure*}

\begin{figure*}
\begin{center}
\includegraphics[width=40pc]{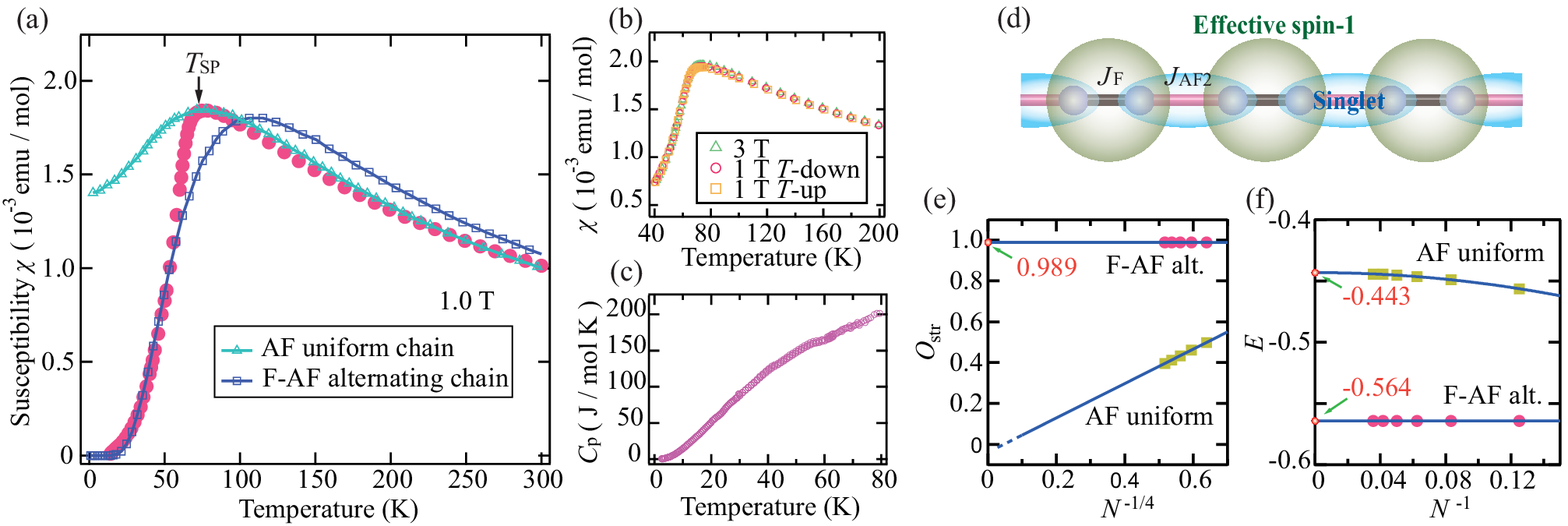}
\caption{(color online) (a) Temperature dependence of magnetic susceptibility ($\chi$ = $M/H$) of ($p$-MePy-V-$p$-CN)PF$_6$$\cdot$CH$_3$CN at 1.0 T, where 3.4 ${\%}$ paramagnetic impurities due to the lattice defects are subtracted assuming the Curie contribution.
The arrow indicates SP transition temperature $T_{\rm{SP}}$.
The solid lines with open triangles and squares represent the results calculated by the QMC method. 
(b) Temperature dependence of $\chi$ near $T_{\rm{SP}}$ at 1.0 T and 3.0 T. $T$-up and $T$-down represent measurements with heating and cooling processes, respectively.
(c) Temperature dependence of the specific heat $C_{\rm{p}}$ of ($p$-MePy-V-$p$-CN)PF$_6$$\cdot$CH$_3$CN at 0 T.
(d) Valence bond picture of the effective Haldane state in the spin-1/2 F-AF alternating chain. 
The ovals represent the valence bond singlet pairs of the two $S$ = 1/2 spins.
(e) System size $N$ dependence of string order parameter $O_{\rm{str}}$ and (f) ground state energy $E$ for the spin-1/2 AF uniform chain and the spin-1/2 F-AF alternating chain. Here, $E$ is normalized by $J_{\rm{AF1}}$ for both cases. The solid lines indicate fitting curves with $N^{-1/4}$ for $O_{\rm{str}}$ and $N^{-2}$ for $E$. The arrows indicate the values evaluated by extrapolation to $N$ $\rightarrow$ $\infty$.
}\label{f2}
\end{center}
\end{figure*}

We observed an SP transition to an effective Haldane state at $T_{\rm{SP}}$ = 70 K in ($p$-MePy-V-$p$-CN)PF$_6$$\cdot$CH$_3$CN, whose molecular structure is shown in Fig.1(a).
The crystallographic parameters at room temperature are as follows: orthorhombic, space group $Pna$2$_1$, $a$ = 7.3758(3) $\rm{\AA}$, $b$ = 15.4561(8) $\rm{\AA}$, $c$ = 21.2673(11) $\rm{\AA}$, V = 2424.5(2) $\rm{\AA}^3$~\cite{supply}.
For $T$ ${\textless}$ $T_{\rm{SP}}$, the space group changed to monoclinic $P$2$_1$ owing to the structural phase transition associated with the SP transition.
The crystallographic parameters at 25 K are as follows: monoclinic, space group $P$2$_1$, $a$ = 7.2203(8) $\rm{\AA}$, $b$ = 20.834(2) $\rm{\AA}$, $c$ = 15.4026(14) $\rm{\AA}$, $\beta$ = 91.776(11)$^{\circ}$, V = 2315.9(4) $\rm{\AA}^3$~\cite{supply}.
Each verdazyl radical $p$-MePy-V-$p$-CN has a spin-1/2, and approximately 62 ${\%}$ of the total spin density is present on the central verdazyl ring (including four N atoms).
The phenyl and cyanophenyl rings account for approximately 15-18 ${\%}$ of the relatively large total spin density, whereas the methylpyridine ring accounts for less than 5 ${\%}$ of the total spin density, yielding the shape of a singly occupied molecular orbital (SOMO), as shown in Fig. 1(b). 
For $T$ ${\textgreater}$ $T_{\rm{SP}}$, the $\pi$-$\pi$ stacking of radicals with glide reflection symmetry forms a 1D uniform structure along the $a$-axis, as shown in Figs. 1(c) and 1(d).
When $T$ ${\textless}$ $T_{\rm{SP}}$, the glide reflection symmetry disappears, and two crystallographically independent molecules form a 1D alternating structure, as shown in Figs. 1(e) and 1(f).
Here, the N-N short contacts in the central verdazyl indicate lattice shrinkage at low temperatures.
Because the nonmagnetic PF$_6$ and CH$_3$CN are located between the 1D chains, the one-dimensionality of the present system is enhanced.
MO calculations were performed to evaluate the exchange interactions between the spins of the molecules forming the 1D chains.
The evaluation presented the following values: $J_{\rm{AF1}}/k_{\rm{B}}$ = 93 K for $T$ ${\textgreater}$ $T_{\rm{SP}}$, $J_{\rm{AF2}}/k_{\rm{B}}$ = 174 K, and $J_{\rm{F}}/k_{\rm{B}}$ = $-$56 K for $T$ ${\textless}$ $T_{\rm{SP}}$; these are defined in the Heisenberg spin Hamiltonian given by $\mathcal {H} = J_{n}{\sum^{}_{<i,j>}}\textbf{{\textit S}}_{i}{\cdot}\textbf{{\textit S}}_{j}$, where $\sum_{<i,j>}$ denotes the sum of the neighboring spin pairs.
The MO calculations indicate that a spin-1/2 AF uniform chain changes to a spin-1/2 F-AF alternating chain at $T_{\rm{SP}}$, as shown in Figs. 1(c) and 1(e).

Figure 2(a) shows the temperature dependence of the magnetic susceptibility ($\chi$ = $M/H$) at 1.0 T.
We observed a broad peak at approximately 74 K, indicating AF correlations in the 1D spin chain.
When the temperature was further decreased, $\chi$ abruptly decreased at $T_{\rm{SP}}$ = 70 K, suggesting the formation of a nonmagnetic singlet state with an excitation gap below $T_{\rm{SP}}$.
We calculated the magnetic susceptibilities of the spin-1/2 Heisenberg AF uniform and F-AF alternating chains using the QMC method, where the ratio $J_{\rm{F}}$/$J_{\rm{AF2}}$ = -0.32, evaluated from the MO calculation, was assumed. 
The experiment and calculation were in good agreement for both temperature regions using the parameters: $J_{\rm{AF1}}/k_{\rm{B}}$ = 119 K, $J_{\rm{AF2}}/k_{\rm{B}}$ = 177 K, and $J_{\rm{F}}/k_{\rm{B}}$ = $-$57 K, as shown in Fig. 2(a). 
The obtained parameters for the spin-1/2 F-AF alternating chain indicate that the ground state has an energy gap of 150 K.
It is confirmed that the evaluation of exchange interactions from MO calculations is reliable in the present case, as in other verdazyl compounds.
At $T_{\rm{SP}}$, temperature hysteresis was not observed in $\chi$, as shown in Fig. 2(b), which is consistent with the characteristics of the second-order SP transition.
In contrast, there was no anomalous behavior associated with the second-order phase transition at $T_{\rm{SP}}$ in specific heat $C_{\rm{p}}$, as shown in Fig. 2(c).
Several organic compounds exhibiting an SP transition did not exhibit phase transition signals at corresponding specific heat values because their crystal structures were not significantly distorted by the SP transitions~\cite{zeroheat1, zeroheat2, zeroheat3}.
The structural change in the present compound was also relatively small, as shown in Figs. 1(c)-1(f).
The phase boundary of a conventional SP system is predicted to have magnetic field dependence~\cite{riron7}.
If we assume the predicted relation, the temperature shift for the present system is expected to be approximately 0.2 K even at 10 T, which is difficult to examine under experimental conditions.
The observed ${\chi}$ at 3 T confirms that the magnetic field dependence of the $T_{\rm{SP}}$ is not significant in the present compound, as shown in Fig. 2(b).

Here, we compare the ground state energies of the AF uniform and F-AF alternating chains assuming the parameters evaluated from the magnetization analysis.
The ground state of the spin-1/2 Heisenberg AF chain is a well-known Tomonaga-Luttinger liquid (TLL), which is a quantum critical state with fermionic spin-1/2 spinon excitation.
In the case of the spin-1/2 Heisenberg F-AF alternating chain, the ground state is essentially the same as the spin-1 Haldane state~\cite{hida1,hida2}.
Two spins coupled by F interaction can be regarded as an effective spin-1, and two spin-1/2 particles on different spin-1 sites form a singlet dimer via AF interaction, as illustrated in Fig. 2(d).
We consider the Heisenberg spin Hamiltonian $H_{\rm{uni}}$ for uniform chain and $H_{\rm{alt}}$ for alternating chain given by 
\begin{multline}
{H_{\rm{uni}}} = {\sum^{N}_{i=1}}(J_{\rm{AF1}}\textbf{{\textit S}}_{i}{\cdot}\textbf{{\textit S}}_{i+i}),\\
{H_{\rm{alt}}} = {\sum^{N}_{i=1}}(J_{\rm{F}}\textbf{{\textit S}}_{2i-1}{\cdot}\textbf{{\textit S}}_{2i}+J_{\rm{AF2}}\textbf{{\textit S}}_{2i}{\cdot}\textbf{{\textit S}}_{2i+1}), 
\end{multline}
where $\textbf{{\textit S}}$ is the spin-1/2 operator, and $N$ is the system size.
The string order parameter for this system is defined by 
\begin{equation}
O_{\rm{str}} =  -4{\langle} {S}^{z}_{2i}{\rm{exp}}[i\pi({S}^{z}_{2i+1}+{S}^{z}_{2i+2}+\cdots+{S}^{z}_{2j-2})]{S}^{z}_{2j-1}{\rangle}, 
\end{equation}
which indicates hidden topological order with a specific value in the Haldane phase~\cite{string1,string2,string3, string4}.
$O_{\rm{str}}$ was evaluated as 0.989 for the present F-AF alternating chain, whereas $O_{\rm{str}}$ for the AF uniform chain approached zero, as shown in Fig. 2(e). 
The evaluated value of $O_{\rm{str}}$ was consistent with the range of 0.38 ${\textless}$$O_{\rm{str}}$${\textless}$1 for the effective spin-1 Haldane phase~\cite{hida1}.
We calculated the ground state energies of the AF uniform and F-AF alternating chains by numerically diagonalizing the Hamiltonian.
Figure 2(f) shows the system size dependence of the calculated ground state energy $E$ per spin site, which was normalized by $J_{\rm{AF1}}$.
It was confirmed that the effective Haldane state for the F-AF alternating chain has a distinctly lower energy than that of the TLL in the AF uniform chain.
If we start from $J_{\rm{F}}$=0, the ground state energy of the alternation chain with $J_{\rm{AF2}}$${\textgreater}$$|J_{\rm{F}}|$ is primarily lowered by the second-order perturbations of $J_{\rm{F}}$; hence, the sign of $J_{\rm{F}}$ does not have a significant effect on the value of $E$.
If we assume the AF-AF alternating chain with a positive ratio of $J_{\rm{F}}$/$J_{\rm{AF2}}$ = 0.32, the ground state energy is evaluated as $E$ = -0.565 at $N$=20, which is extremely close to $E$ = -0.564 of the F-AF alternating chain.
Moreover, the $E$ for the AF dimer ($J_{\rm{F}}$ =0) given by -(3/4)($J_{\rm{AF2}}$/$J_{\rm{AF1}}$)/2 = -0.558 is also close to that of the alternating chains, demonstrating that the most critical factor that lowers the ground state energy (gain of magnetic energy) in the SP transition is the absolute value of $J_{\rm{AF2}}$.

\begin{figure}
\begin{center}
\includegraphics[width=21pc]{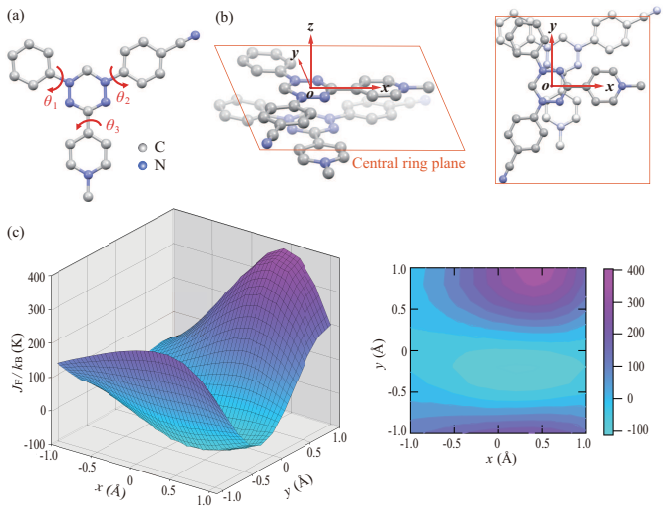}
\caption{(color online) (a) Three degrees of freedom for dihedral angles in $p$-MePy-V-$p$-CN.
(b) Coordinate system with the origin as the average position of four nitrogen atoms for molecular pairs associated with the exchange interactions. 
The $x$-axis is defined in the direction of the methylpyridine ring, and the $xy$-plane is defined to be parallel to the central ring of the molecule.
(c) Change of $J_{\rm{F}}$ in the $xy$-plane simulated by the MO calculation. 
The origin is defined at the position of the actual molecular pair associated with $J_{\rm{F}}$.
}\label{f3}
\end{center}
\end{figure}

Here, we examine the conversion of the AF uniform chain to the F-AF alternating chain, that is, the conversion of $J_{\rm{AF1}}$ to $J_{\rm{F}}$ and $J_{\rm{AF2}}$, in terms of molecular orbital coupling.
Our verdazyl radical can exhibit a delocalized $\pi$-electron spin density even on non-planar molecular structures, yielding flexible MOs, which allows the modulation of intermolecular exchange interactions.
The MO of the $p$-MePy-V-$p$-CN radical was modulated by changing the dihedral angle at the SP transition.
The dihedral angles ${\theta}_{1}$, ${\theta}_{2}$, and ${\theta}_{3}$ in Fig. 3(a) for the two crystallographically independent molecules in the low-temperature phase exhibited changes from high-temperature phase of ($\Delta{\theta}_{1}$, $\Delta{\theta}_{2}$, $\Delta{\theta}_{3}$)=(3.0$^{\circ}$, -4.4$^{\circ}$, 0.9$^{\circ}$) and (1.6$^{\circ}$, 7.2$^{\circ}$, -4.7$^{\circ}$), respectively.
Moreover, to evaluate the changes in intermolecular distance associated with the SP transition, we defined a coordinate system with the origin as the average position of four nitrogen atoms, as shown in Fig. 3(b). 
As the $xy$-plane is defined to be parallel to the central ring of the molecule, the changes in the $z$-direction and the $xy$-plane almost correspond to the intermolecular distance and the lateral shift of the facing molecular pairs associated with the exchange interactions.
Accordingly, we evaluated the changes in the values mentioned above from the position in the high-temperature phase: (${\Delta}x$, ${\Delta}y$, ${\Delta}z$) = (0.03 $\rm{\AA}$, -0.03 $\rm{\AA}$, -0.14 $\rm{\AA}$) for $J_{\rm{AF2}}$ and (${\Delta}x$, ${\Delta}y$, ${\Delta}z$) = (0.07 $\rm{\AA}$, 0.26 $\rm{\AA}$, -0.16 $\rm{\AA}$) for $J_{\rm{F}}$.
For $J_{\rm{AF2}}$, the lateral shift was very slight, revealing that the main structural change accompanying the SP transition was a contraction in the 1D stacking direction. 
The evaluated ${\Delta}z$ was actually identical to the change in the N-N short contact.
Conversely, a relatively large lateral shift for the molecular pair associated with $J_{\rm{F}}$ was observed.
Lateral shifts often reduce the energy gap between the highest occupied MO (HOMO) and the lowest unoccupied MO (LUMO) and produce mostly degenerate HOMO and LUMO, which leads to intermolecular F exchange interactions.
We simulated the change in $J_{\rm{F}}$ with respect to the lateral shift using the MO calculation, in which the origin is defined at the actual position of the molecular pair associated with the $J_{\rm{F}}$, as shown in Fig. 3(c).
The sign of $J_{\rm{F}}$ changed dramatically depending on the lateral shift. 
In particular, the changes in the $y$-direction were remarkable.
Considering that the largest shift of 0.26 $\rm{\AA}$ is estimated in the $y$-direction, the calculated data demonstrate that the drastic change in the intermolecular exchange interaction can be caused by the structural change associated with the SP transition.
The face-to-face approach observed in the molecular pair associated with $J_{\rm{AF2}}$ enhances the overlapping of SOMOs, which generally increases the AF exchange interactions. 
Given that the ground state energy of the alternation chain strongly depends on the magnitude of the AF interaction, as described in the numerical analysis, the present system gains magnetic energy by increasing the AF interaction, that is, the conversion of $J_{\rm{AF1}}$ to $J_{\rm{AF2}}$.
Therefore, the lattice distortion that enhance the AF interaction at the SP transition is considered to induce the lateral shift of another molecular pair in the 1D stacking contributing to a stable MO overlap, resulting in the conversion of $J_{\rm{AF1}}$ to $J_{\rm{F}}$.

The conventional SP transition to the AF-AF alternating chain predicted $T_{\rm{SP}}=0.8{\eta}J/k_{\rm{B}}$, where $J$ is the exchange interaction in the uniform chain, and ${\eta}$ is the generalized spin-lattice coupling parameter~\cite{riron6}.
If this relationship is applied to the present system, we obtain ${\eta}$ $\approx$ 0.74.
The SP systems reported thus far, including inorganic compounds, have smaller ${\eta}$ values.
The large value of ${\eta}$ in the present system suggests that the flexibility of MOs in our radical systems yields effective strong spin-orbit coupling.


In summary, we successfully synthesized a model compound that exhibited an unconventional SP transition.
The spin-1/2 uniform AF chain was converted to a spin-1/2 F-AF alternating chain, forming the effective spin-1 Haldane state at the SP transition.
The present results demonstrate the flexibility of MOs in organic radical systems that can easily change the magnitude and sign of exchange interactions, realizing the unconventional quantum phenomena caused by spin-lattice couplings.
In this study, we provide a research area focusing on quantum phenomena generated by strong spin-lattice couplings in radical-based magnets.

\begin{acknowledgments}
We thank S. Shimono and Y. Kubota for valuable discussions and Y. Hosokoshi for letting us use the laboratory equipments.
This research was partly supported by the Asahi Glass Foundation and the joint-research program of the Institute for Molecular Science.
\end{acknowledgments}



\begin{thebibliography}{99}
\bibitem{SP1}
E. Pytte, Phys. Rev. B \textbf{10}, 4637 (1974).

\bibitem{SP2}
L. N. Bulaevskii, A. I. Buzdin, and D. I. Khomskii, Solid State Communications \textbf{27}, 5 (1978).

\bibitem{Peierls}
R. E. Peierls, Quantum Theory of Solids (Oxford University Press, Oxford, UK, 1955).

\bibitem{SP_exp1}
J. W. Bray, H. R. Hart, L. V. Interrante, I. S. Jacobs, J. S. Kasper, G. D. Watkins, S. H. Wee, and J. C. Bonner, Phys. Rev. Lett. \textbf{35}, 744 (1975).

\bibitem{SP_exp2}
S. I. Jacobs, J. W. Bray, H. R. Hart, L. V. Interrante, J. S. Kasper, G. D. Watkins, D. E. Prober, and J. C. Bonner, Phys. Rev. B \textbf{14}, 3036 (1976).

\bibitem{SP_exp3}
M. Hase, I. Terasaki, and K. Uchinokura, Phys. Rev. Lett. \textbf{70}, 3651 (1993).

\bibitem{SP_exp4}
D. S. Chow, P. Wzietek, D. Fogliatti, B. Alavi, D. J. Tantillo, C. A. Merlic, and S. E. Brow, Phys. Rev. Lett. \textbf{81}, 3984 (1998).

\bibitem{SP_exp5}
W. Fujita, K. Awaga, R. Kondo, S. Kagoshima, J. Am. Chem. Soc. \textbf{128}, 6016 (2006).

\bibitem{SP_exp6}
K. Mukai, N. Wada, J. B. Jamali, N. Achiwa, Y. Narumi, K. Kindo, T. Kobayashi, and K. Amaya, Chem. Phys. Lett. \textbf{257}, 538 (1996).

\bibitem{SPtoSC_01}
D. J$\rm{\acute{e}}$rome, Science \textbf{252}, 1509 (1991).

\bibitem{SPtoSC_02}
T. Adachi, E. Ojima, K. Kato, H. Kobayashi, T. Miyazaki, M. Tokumoto, and A. Kobayashi, J. Am. Chem. Soc. \textbf{122}, 3238 (2000).

\bibitem{SPtoSC_03}
D. Jaccard, H. Wilhelm, D. J$\rm{\acute{e}}$rome, J. Moser, C. Carcel, and J. M. Fabre, J. Phys. Condens. Matter \textbf{13}, L89 (2001)

\bibitem{SPtoSC_04}
H. Mori, J. Phys. Soc. Jpn. \textbf{75}, 051003 (2006).

\bibitem{haldane}
F. D. M. Haldane, Phys. Rev. Lett. \textbf{50}, 1153 (1983).

\bibitem{AKLT}
I. Affleck, T. Kennedy, E. H. Lieb, and H. Tasaki, Phys. Rev. Lett. \textbf{59}, 799 (1987).

\bibitem{qcom1}
A. Miyake, Phys. Rev. Lett. \textbf{105}, 040501 (2010).

\bibitem{qcom2}
S. D. Bartlett, G. K. Brennen, A. Miyake, and J. M. Renes, Phys. Rev. Lett. \textbf{105}, 110502 (2010).

\bibitem{qcom3}
D. V. Else, I. Schwarz, S. D. Bartlett, and A. C. Doherty, Phys. Rev. Lett. \textbf{108}, 240505 (2012).


\bibitem{CuNbO_2}
K. Kodama, H. Harashina, H. Sasaki, M. Kato, M. Sato, K. Kakurai, and M. Nishi, J. Phys. Soc. Jpn. \textbf{68}, 237 (1999).

\bibitem{IPACu_masuda}
T. Masuda, A. Zheludev, H. Manaka, L.-P. Regnault, J.-H. Chung, and K. Qiu, Phys. Rev. Lett. \textbf{96}, 047210 (2006).

\bibitem{DMA_neutron}
M. B. Stone, W. Tian, M. D. Lumsden, G. E. Granroth, D. Mandrus, J.-H. Chung, N. Harrison, and S. E. Nagler, Phys. Rev. Lett. \textbf{99}, 087204 (2007).

\bibitem{NaCuSbO_2}
Y. Miura, Y. Yasui, T. Moyoshi, M. Sato, and K. Kakurai, J. Phys. Soc. Jpn. \textbf{77}, 104709 (2008).

\bibitem{Li3Cu2SbO6}
A. Bhattacharyya, T. K. Bhowmik, D. T. Adroja, B. Rahaman, S. Kar, S. Das, T. Saha-Dasgupta, P. K. Biswas, T. P. Sinha, R. A. Ewings, D. D. Khalyavin, and A. M. Strydom, Phys. Rev. B \textbf{103}, 174423 (2021).

\bibitem{hida1}
K. Hida, Phys. Rev. B \textbf{45}, 2207 (1992).

\bibitem{hida2}
K. Hida, J. Phys. Soc. Jpn. \textbf{62}, 1463 (1993).

\bibitem{3Cl4FV} 
H. Yamaguchi, K. Iwase, T. Ono, T. Shimokawa, H. Nakano, Y. Shimura, N. Kase, S. Kittaka, T. Sakakibara, T. Kawakami, and Y. Hosokoshi, Phys. Rev. Lett. {\bf 110}, 157205 (2013).

\bibitem{a26Cl2V} 
H. Yamaguchi, T. Okubo, S. Kittaka, T. Sakakibara, K. Araki, K. Iwase, N. Amaya, T. Ono, and Y. Hosokoshi, Sci. Rep. {\bf 5}, 15327 (2015).

\bibitem{random}
H. Yamaguchi, M. Okada, Y. Kono, S. Kittaka, T. Sakakibara, T. Okabe, Y. Iwasaki, and Y. Hosokoshi, Sci. Rep. {\bf 7}, 16144 (2017).


\bibitem{fine-tune} 
H. Yamaguchi, H. Miyagai, T. Shimokawa, K. Iwase, T. Ono, Y. Kono, N. Kase, K. Araki, S. Kittaka, T. Sakakibara, T. Kawakami, K. Okunishi, and Y. Hosokoshi, J. Phys. Soc. Jpn. {\bf 83}, 033707 (2014).


\bibitem{F-AF}
H. Yamaguchi, Y. Shinpuku, T. Shimokawa, K. Iwase, T. Ono, Y. Kono, S. Kittaka, T. Sakakibara, and Y. Hosokoshi, Phys. Rev. B {\bf 91}, 085117 (2015).

\bibitem{square_PF6}
H. Yamaguchi, Y. Sasaki, T. Okubo, M. Yoshida, T. Kida, M. Hagiwara, Y. Kono, S. Kittaka, T. Sakakibara, M. Takigawa, Y. Iwasaki, and Y. Hosokoshi, Phys. Rev. B {\bf 98}, 094402 (2018).

\bibitem{square_SbF6}
H. Yamaguchi, Y. Iwasaki, Y. Kono, T. Okubo, S. Miyamoto, Y. Hosokoshi, A. Matsuo, T. Sakakibara, T. Kida, and M. Hagiwara, Phys. Rev. B, \textbf{103}, L220407 (2021).


\bibitem{procedure}
R. Kuhn, Angew. Chem. \textbf{76}, 691 (1964).

\bibitem{mukai}
K. Mukai, S. Jinno, Y. Shimobe, N. Azuma, M. Taniguchi, Y. Misaki, K. Tanaka, K. Inoue, and Y. Hosokoshi, J. Mater. Chem. \textbf{13}, 1614 (2003).

\bibitem{MOcal} 
M. Shoji, K. Koizumi, Y. Kitagawa, T. Kawakami, S. Yamanaka, M. Okumura, and K. Yamaguchi, Chem. Phys. Lett. {\bf 432}, 343 (2006).

\bibitem{QMC33} 
A. W. Sandvik, Phys. Rev. B {\bf 59}, 14157 (1999).

\bibitem{QMC34} 
A. F. Albuquerque, F. Alet, P. Corboz, P. Dayal, A. Feiguin, L. Gamper, E. Gull, S. Gurtler, A. Honecker, R. Igarashi, M. Korner, A. Kozhevnikov, A. Lauchli, S. R. Manmana, M. Matsumoto, I. P. McCulloch, F. Michel, R. M. Noack, G. Pawlowski, L. Pollet, T. Pruschke, U. Schollwock, S. Todo, S. Trebst, M. Troyer, P. Werner, and S. Wessel, J. Magn. Magn. Mater. {\bf 310}, 1187 (2007)

\bibitem{QMC35} 
B. Bauer, L. D. Carr, A. Feiguin, J. Freire, S. Fuchs, L. Gamper, J. Gukelberger, E. Gull, S. Guertler, A. Hehn, R. Igarashi, S.V. Isakov, D. Koop, P.N. Ma, P. Mates, H. Matsuo, O. Parcollet, G. Pawlowski, J.D. Picon, L. Pollet, E. Santos, V.W. Scarola, U. Schollw$\ddot{\rm{o}}$ck, C. Silva, B. Surer, S. Todo, S. Trebst, M. Troyer, M.L. Wall, P. Werner, and S. Wessel, J. Stat. Mech.: Theory and Experiment, P05001 (2011).

\bibitem{supply} 
See Supplemental Material at ... for detailed crystallographic parameters.

\bibitem{zeroheat1}
K. Mukai, M. Yanagimoyo, S. Tanaka, M. Mito, T. Kawae, and K. Takeda, J. Phys. Soc. Jpn. {\bf 72}, 2312 (2003).

\bibitem{zeroheat2}
Y. Yoshida, O. Wada, Y. Nakaie, T. Kawae, N. Sakai, N. Kawame, Y. Fujii, Y. Hosokoshi, B. Grenier, and J-P. Boucher, J. Mag. Mag. Mater. {\bf 310}, 1215 (2007).

\bibitem{zeroheat3}
G. Guan, S. Fukuoka, S. Yamashita, T. Yamamoto, H. Taniguchi, and Y. Nakazawa, J. Therm. Anal. Cal. {\bf 113}, 1197 (2013).

\bibitem{riron7}
M. C. Cross, Phys. Rev. B {\bf 20}, 4606 (1979).

\bibitem{string1}
M. den Nijs and K. Rommelse, Phys. Rev. B \textbf{40}, 4709 (1989).

\bibitem{string2}
H. Tasaki, Phys. Rev. Lett. \textbf{66}, 798 (1991).

\bibitem{string3}
S. M. Girvin and D. P. Arovas, Phys. Scr. \textbf{T27}, 156 (1989).

\bibitem{string4}
S. R. White and D. A. Huse, Phys. Rev. B \textbf{48}, 3844 (1993).

\bibitem{riron6}
M. C. Cross and D. S. Fisher, Phys. Rev. B {\bf 19}, 402 (1979).


\end{thebibliography}
\end{document}